\documentclass{ptephy_v1}%%%%%% to generate preprint number with ptep logo

\preprintnumber{XXXX-XXXX} %%% %%% Insert preprint number here

\usepackage{xcolor}

\begin{document}

\title{Characterization of two-level system noise for microwave kinetic inductance detector comprising niobium film on silicon substrate}

\author{Y. Sueno}
\affil{Department of Physics, Graduate School of Science, Kyoto University, Kitashirakawa-Oiwakecho, Sakyo-ku, Kyoto 606-8502, Japan \email{sueno.yoshinori.83x@st.kyoto-u.ac.jp}}

\author{S. Honda}
\affil{Instituto de Astrofisica de Canarias
Calle Vía Láctea, s/n, 38205 San Cristóbal de La Laguna, Santa Cruz de Tenerife, Spain}

\author{H. Kutsuma}
\affil{RIKEN Center for Quantum Computing (RQC), RIKEN
2-1 Hirosawa, Wako, Saitama, 351-0198, Japan}

\author{S. Mima}
\affil{Superconductive ICT Device Laboratory, Kobe Frontier Research Center, Advanced ICT Research Institute, National Institute of Information and Communications Technology
588-2, Iwaoka, Nishi-ku, Kobe, Hyogo, 651-2492, Japan}

\author{C. Otani}
\affil{Terahertz Sensing and Imaging Research Team, RIKEN 519-1399 Aramaki-Aoba, Aoba-ku, Sendai, 980-0845, Miyagi, Japan}

\author{S. Oguri}
\affil{Institute of Space and Astronautical Science (ISAS), Japan Aerospace Exploration Agency (JAXA)
3-1-1 Yoshinodai, Chuo-ku, Sagamihara, Kanagawa 252-5210, Japan}

\author[1]{J. Suzuki}

\author[1]{O. Tajima}

\begin{abstract}
A microwave kinetic inductance detector (MKID) is a cutting-edge superconducting detector. It comprises a resonator circuit constructed with a superconducting film on a dielectric substrate.
To expand its field of application, it is important to establish a method to suppress the two-level system (TLS) noise
that is caused by the electric fluctuations between the two energy states at the surface of the substrate.
The electric field density can be decreased by expanding the strip width (S) and gap width from the ground plane (W) in the MKID circuit,
allowing the suppression of TLS noise.
However, this effect has not yet been confirmed for MKIDs made with niobium films on silicon substrates.
In this study, we demonstrate its effectiveness for such MKIDs.
We expanded the dimension of the circuit from (S, W) = (3.00 $\mu$m, 4.00 $\mu$m) to (S, W) = (5.00 $\mu$m, 23.7 $\mu$m), and achieved an increased suppression of 5.5 dB in TLS noise.

\end{abstract}

\subjectindex{xxxx, xxx}

\maketitle

\section{Introduction}

A microwave kinetic inductance detector (MKID) is a cutting-edge superconducting detector~\cite{1,2}.
Its readout scheme is based on natural frequency-domain multiplexing. 
This enables the reading of a large number of detector signals by using a pair of readout cables, ${\mathcal{O}(1000)}$~\cite{3,4}.
It thus enables the minimization of the heat load entering the detectors via cables. 
These advantages ensure that MKIDs are suitable for applications in radio astronomy~\cite{5,6,7}.

\begin{figure}[tb!]
  \centering
  \includegraphics[width = 15cm]{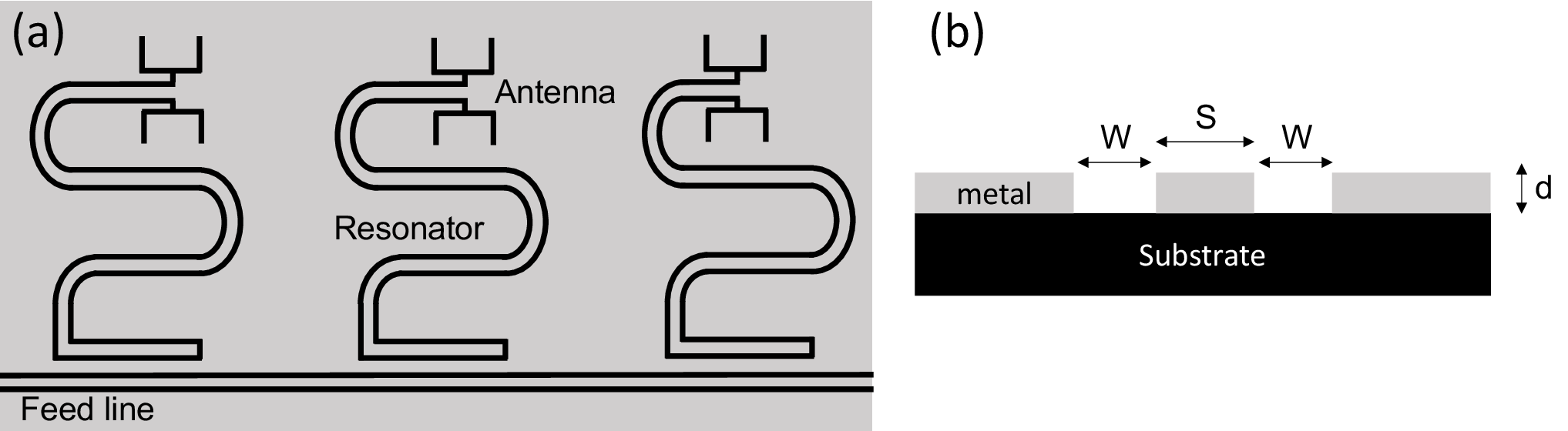}
  \caption{(a) Schematic of the MKIDs. Each MKID comprises an antenna-coupled superconducting resonator that is coupled to the feed line.
  (b) Cross-sectional view of the CPW for the resonator film. In this study, we define the center strip width, gap width, and thickness as S, W, and d, respectively.}
  \label{wide_pic}
\end{figure}

For the radio detection, each MKID comprises an antenna-coupled superconducting resonator that is coupled to a feed line.
As illustrated in Figure~\ref{wide_pic}, the resonator circuit and feed line are constructed using a coplanar waveguide (CPW) on a dielectric substrate.
Resonances of each MKID are monitored with the transmittance of the microwaves fed into the MKID.
The radiation entering via the antenna breaks the Cooper pairs in the resonator film.\footnote{We phrase Cooper-pair breaking as quasiparticle generation.}
This is detected as a variation in the resonance, i.e., a variation in the resonant depth (known as amplitude response),
and
a variation in the resonant phase (known as phase response).
The phase response is more commonly used because its responsivity is approximately one order of magnitude higher than that of the amplitude response~\cite{8,9}.

\begin{figure}[tb!]
  \centering
  \includegraphics[width = 10cm]{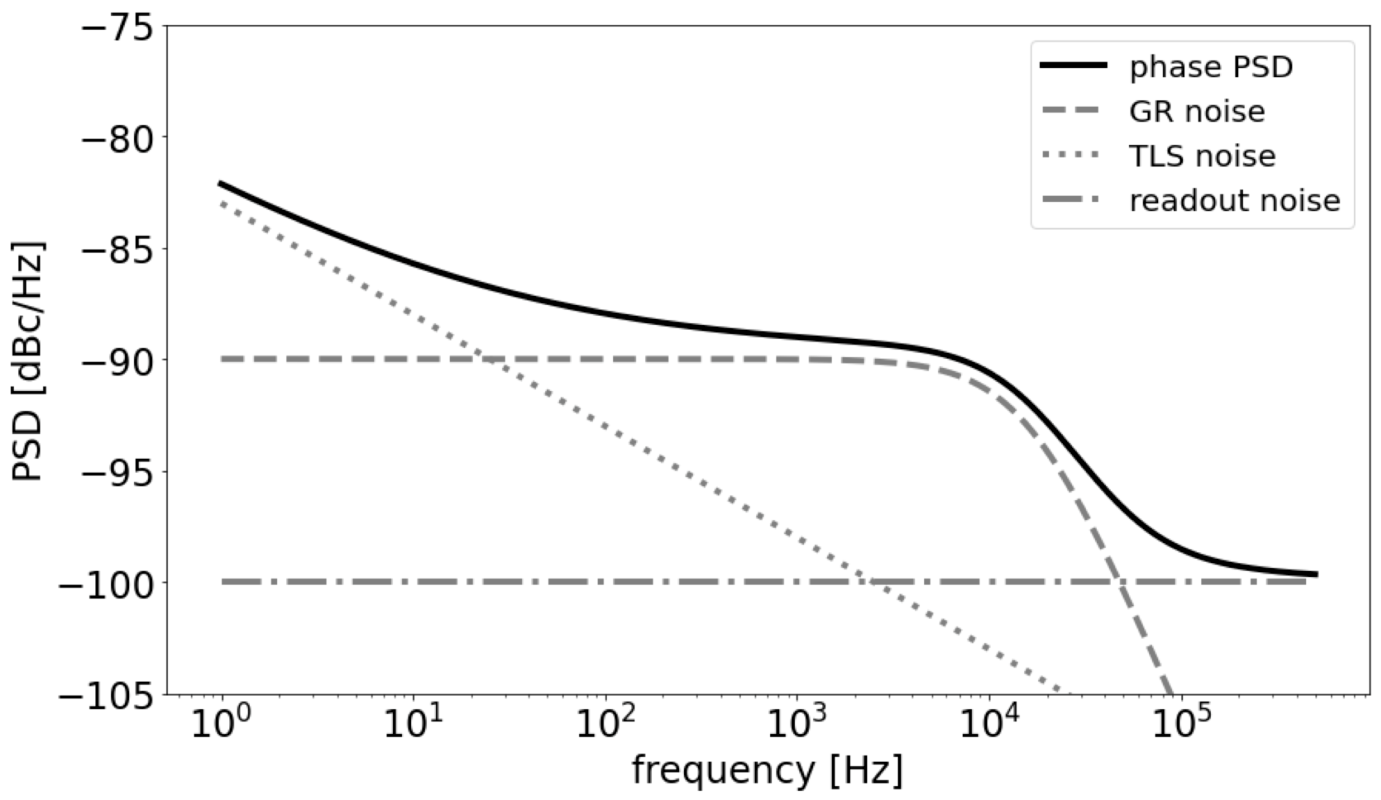}
  \caption{Illustration of typical noise spectrum of the MKID in the phase response.
  It consists of three noise components: 
  GR noise, TLS noise, and readout noise.
  }
  \label{PSD_noise}
\end{figure}

The noise property of the MKID is usually characterized in a Fourier space.
In Figure~\ref{PSD_noise}, we illustrate the typical shape of a power spectrum density~(PSD) in the phase response.
There are three noise components: generation-recombination (GR) noise, two-level system (TLS) noise, and readout noise~\cite{10,11}.
A source of GR noise is the generation and recombination of the quasiparticles in the resonator.
Moreover, it is proportional to the number of quasiparticles, which depends on the physical temperature of the MKID and the critical temperature ($T_c$) of the film material, as well as the power of the input signals via the antenna. 
The mean time for the recombination, known as quasiparticle life time~($\tau_{\rm qp}$), is typically 10~$\mu$s -- 1,000~$\mu$s
in the case of MKIDs made with an aluminum film. 
Therefore, there is no MKID response above this frequency, and the GR noise behaves as white noise below the frequency of $1/(2 \pi \tau_{\rm qp})$.
The readout noise is dominated by the thermal noise of a cold low-noise amplifier (C-LNA) in the readout chain.
This is generally maintained below the level of the GR noise.
As described in the next section, the TLS noise is caused by electrical fluctuations at the surface of the substrate.
This noise is frequency-dependent, and is of a higher magnitude in the lower frequency region.
Therefore, it is difficult to maintain a low noise in the low-frequency region. 
Thus,
the mitigation of TLS noise can enable the extension of the low noise to the low-frequency region.

The magnitude of the TLS noise depends on the superconducting material at the resonator and the substrate material.
Concerning the effect of substrate materials, a previous study demonstrated that an MKID using a silicon substrate achieved lower levels of TLS noise than an MKID using a sapphire substrate~\cite{12}.
Similarly, studies conducted to assess the effect of superconducting materials demonstrated that aluminum generates higher levels of TLS noise than other materials~\cite{13}.

However, aluminum ensures sensitivity to millimeter or sub-millimeter waves owing to its low gap energy that is necessary for Cooper-pair breaking.
Therefore, minimizing the amount of aluminum used in the MKID circuit is a practical approach to mitigate the TLS noise, based on which a hybrid-type MKID~\cite{14} was designed.
Its resonator comprises two different materials, and several combinations can be considered: niobium titanium nitride (NbTiN) and aluminum, titanium nitride (TiN) and aluminum, or niobium and aluminum.
Although this hybrid-type MKID using NbTiN or TiN achieved good TLS noise suppression~\cite{15}, the facilities that can use these materials are limited. 
Moreover, their fabrication is not simple~\cite{16}.
Therefore, it is beneficial to develop a methodology of TLS noise reduction for niobium MKID.

Expanding the CPW geometry is another approach to suppress the TLS noise~\cite{17}.
To assess this approach, 
a study was conducted on MKIDs using niobium films on sapphire substrates~\cite{17}.
However, there have been no published studies on MKIDs using niobium films on silicon substrates. 

In this study, we characterized the TLS noise for an MKID using a niobium film on a silicon substrate, and compared the magnitude of the TLS noise between two different CPW geometries.
Our methods and results are presented as follows:
In section 2, we describe the properties of TLS noise.
In section 3, we describe the design and fabrication of the MKID devices.
In section 4, we characterize the TLS noise of these MKIDs.
In section 5, we discuss the difference in TLS noise between the two different CPW geometries.
Finally, we state our conclusions in section 6.

\section{Two-level system (TLS) noise}

\begin{figure}[tb!]
  \centering
  \includegraphics[width = 15cm]{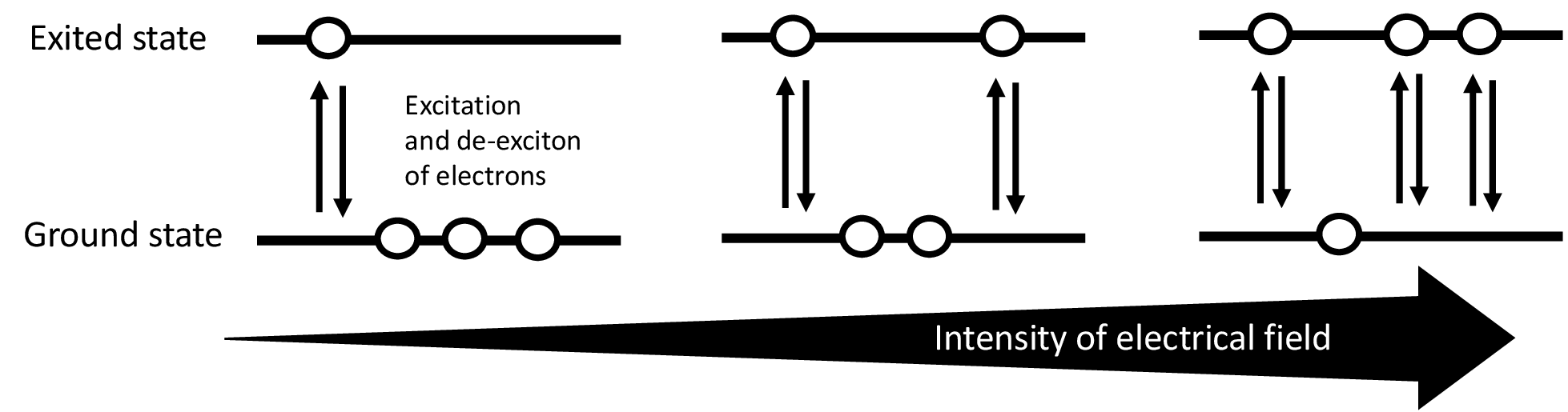}
  \caption{Mechanism of TLS fluctuation. The electrical field of the readout microwaves excites the energy states from the ground state to the excited state.
  De-excitation occurs over a finite length of time.
  These excitations and de-excitations are the origins of the electrical fluctuations.
  Therefore, the TLS noise depends on the frequency and power of the readout microwaves.}
  \label{TLS_effect}
\end{figure}

TLS noise is caused by electrical fluctuations between the ground state and an excited energy state at the surface of the substrate.
The fluctuation between the two energy states are caused by readout microwaves that create electric fields at the surface of the CPW, as illustrated in Figure \ref{TLS_effect}.
As a result, fluctuations in the dielectric constant of the substrate cause fluctuations in the resonant frequency of the MKID. 
TLS noise can be characterized based on its dependences on three parameters. First, it has frequency dependence owing to the time intervals from excitation to de-excitation between the two energy states.
Second, it also has the readout power dependence which can be characterized by using the power law to the readout power.
\footnote{In general, the lower noise level is obtained with the higher readout power. However, the TLS fluctuation also depends on the readout power. We observe a combination of them in the PSD.}
Third, the TLS noise also has temperature dependence.
The occupancy of the excited states and the relaxation rate tend to be high when the device temperature is high~\cite{18}.
Therefore, we obtain lower TLS noise at higher device temperatures.

\begin{figure}[tb!]
  \centering
  \includegraphics[width = 12cm]{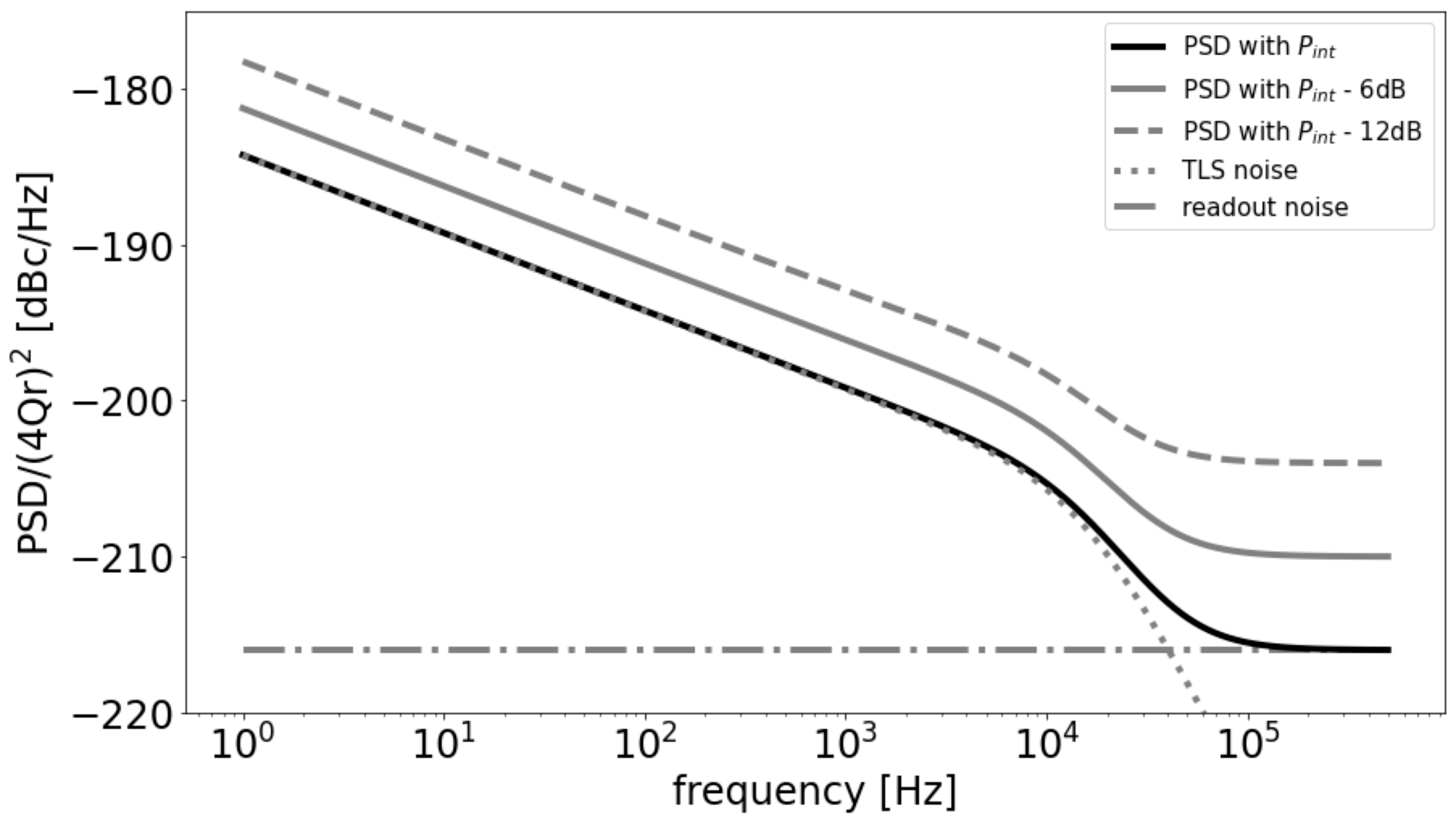}
  \caption{Illustration of the expected PSDs for the MKID using the niobium on the silicon substrate.
  The vertical axis is normalized by $(4Q_r)^2$ for the TLS noise study, as represented in Eq. (\ref{eq:PSD_TLS}). Here, we assume $Q_r = 10^5$.
  For the MKID fully made with the niobium film, the GR noise is negligible because its $T_c$ is high (9.2 K).
  The response speed of the readout microwaves in the resonator is described as the resonator ring time ($\tau_{\rm res} = Q_r/\pi f_r$), which is typically $0.1 \mu s - 10 \mu s$. There is no response to TLS fluctuation above the frequency of $1/(2 \pi \tau_{\rm res}$).}
  \label{tls_expect}
\end{figure}

We modeled the PSD for the TLS noise study based on the frequency ($f$), 
stored internal power in the resonator circuit ($P_{\rm int}$),
and device temperature ($T$),
\begin{equation}
  PSD =  (4 Q_r)^2 \times N_{\rm TLS} \left(f \over 475~{\rm Hz}\right)^{\alpha_f} \left( P_{\rm int} \over -40~{\rm dBm}\right)^{\alpha_P}  \left( { T \over 0.33~{\rm K}} \right)^{\alpha_T} ,
  \label{eq:PSD_TLS}
\end{equation}
where $Q_r$ is a quality factor of the resonator; 
$N_{\rm TLS}$ represents the magnitude of the TLS noise; and $\alpha_f, \alpha_P$, and $\alpha_T$ are the power-law indices of frequency, internal power, and device temperature, respectively.
The $P_{\rm int}$ was calculated using the following formula according to the reference~\cite{18},
\begin{equation}
  P_{\rm int} = \frac{2 Q_r^2}{\pi Q_c} P_{\rm read}, 
\end{equation}
where $Q_c$ is a quality factor related to the coupling between the resonator and the feed line, and $P_{\rm read}$ is the power of the readout microwaves in the feed line.
Then, we measured PSDs by changing the readout power as illustrated in Figure~\ref{tls_expect}. 
Following this, we extracted the model parameters, $N_{\rm TLS}$, $\alpha_f$, and $\alpha_P$ from the measured PSDs.
As discussed above, we expected to negative values for $\alpha_f$, $\alpha_P$ and $\alpha_T$.
For the MKID made with the NbTiN film on a silicon substrate,
these parameters were measured at
$\alpha_f \approx -0.5$, $\alpha_P \approx -0.5$, and $\alpha_T \approx -2$, respectively~\cite{19};
identical results were obtained for the MKID with niobium film on a sapphire substrate~\cite{20}.
In this study, we assumed $\alpha_T = -2$ considering the limited cooling capacity.

\begin{figure}[t!]
  \centering
  \includegraphics[width = 12cm]{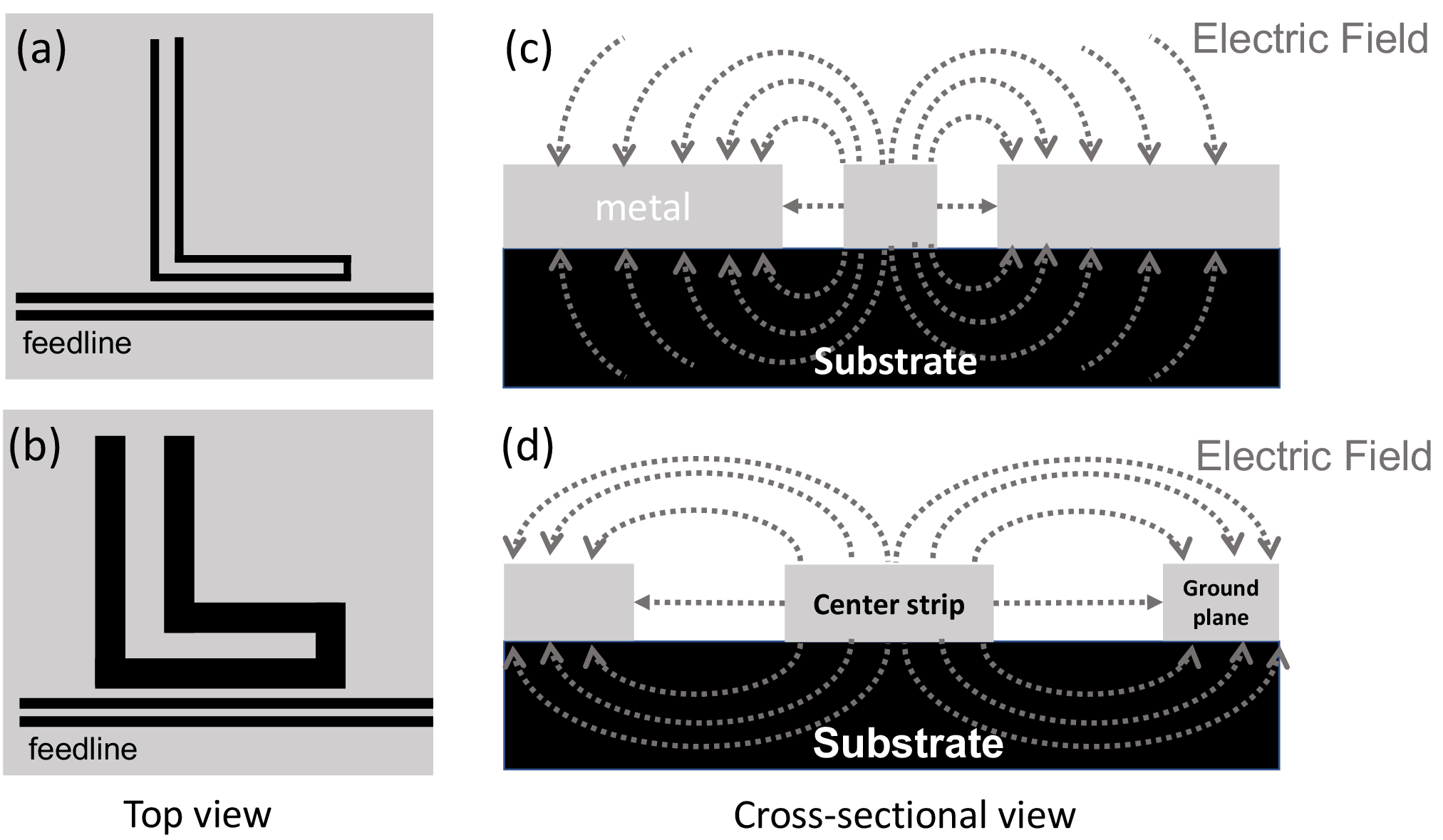}
  \caption{Electric field density depends on the CPW geometry.
  Top and cross-sectional views of the CPW for MKIDs with narrow (a,c) and wide (b,d) geometries.
  Electric fields from the center strip line to the ground plane are illustrated in the cross-sectional views.
  The wide geometry has a lower electric field density than the narrow geometry.
  }
  \label{TLS_geo_sketch}
\end{figure}

The $N_{\rm TLS}$ depends on the electric field density.
We expected a lower electric field density with a wider CPW structure, as illustrated in Figure~\ref{TLS_geo_sketch}.
The concept we demonstrated in this study is that a lower $N_{\rm TLS}$ is obtained with a wider CPW structure. We expected to obtain approximately four times lower than $N_{\rm TLS}$ with four times CPW structure.

\section{Design and fabrication of MKIDs}
We fabricated two types of MKIDs: one with wide CPW geometry and the other with narrow CPW geometry.
The main differences in their dimensions are the strip width and gap width in the CPW, as summarized in Table~\ref{geo_comp}.
The MKID with wide CPW geometry was denoted as ``Wide-MKID'' and the other as ``Narrow-MKID''. 
There were simulation studies for relations between the electric field density and the CPW geometries~\cite{17, Wang}, and they roughly obtained that the electric field density is inversely proportional to the strip width and gap width. According to this knowledge, we expect that the electric field density of the Wide-MKID would be approximately 6 dB (i.e., four times) lower than that of the Narrow-MKID.
Hence, the Wide-MKID generate lower TLS noise than the Narrow-MKID.
%According to numerical calculations, the electric field density is inversely proportional to the sum of the strip width and gap width.
%Therefore, we estimated that the electric field density of the Wide-MKID would be approximately 6 dB (i.e., four times) lower than that of the Narrow-MKID.
%Hence, the Wide-MKID generate lower TLS noise than the Narrow-MKID.

\begin{figure}[b!]
  \centering
  \includegraphics[width = 15cm]{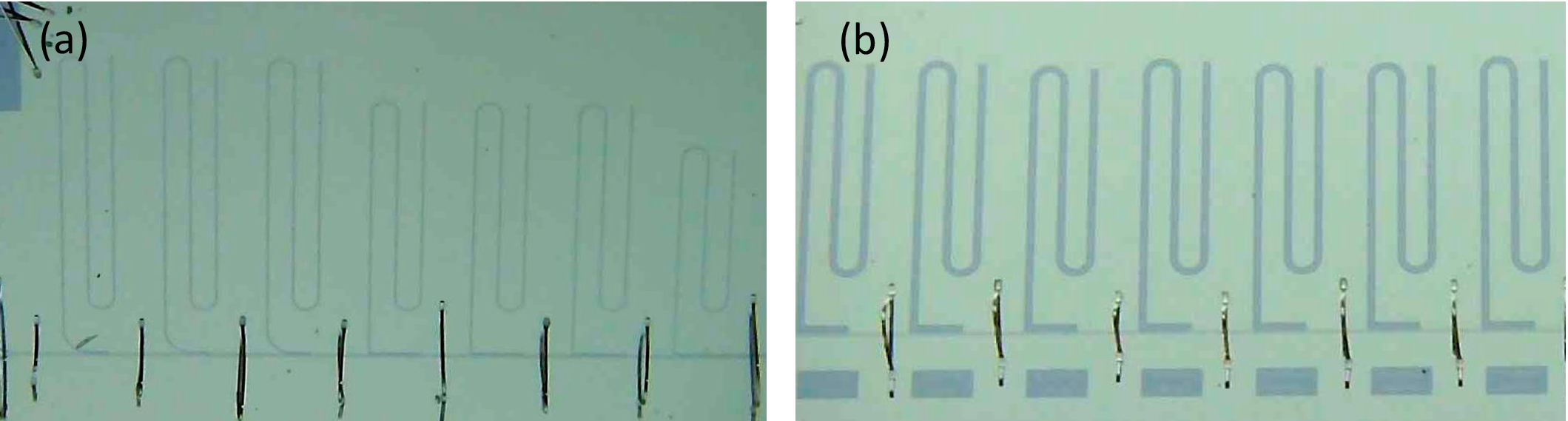}
  \caption{Images of the Narrow-MKID (a), the Wide-MKID (b).}
  \label{fig:photo}
\end{figure}

The MKIDs used in this study were fabricated using equipment at RIKEN.
They comprised a niobium film of 200 nm thickness on a ⟨100⟩-oriented Si wafer.
Figure~\ref{fig:photo} presents their images.
In the next section, we characterized eight Wide-MKIDs and nine Narrow-MKIDs.
\begin{table}[b!]
  \caption{CPW parameters for two types of MKIDs used in this study.
  }
  \label{geo_comp}
  \centering
  \scalebox{0.9}{ %%% scale
  \begin{tabular}{l c c c c c}
  \hline \hline
  & Substrate & Film material & Film thickness (d) & Strip width (S) & Gap width (W) \\
  \hline
  Narrow-MKID & silicon & niobium & 200~nm & 3.00~${\rm \mu m}$ & 4.00~${\rm \mu m}$ \\
  Wide-MKID & silicon & niobium & 200~nm & 5.00~${\rm \mu m}$ & 23.7~${\rm \mu m}$\\
  \hline \hline
  \end{tabular}
  }
\end{table}
We note that the MKIDs used in this study were designed specifically for the TLS noise study and are not hybrid-type MKIDs.
They only use the niobium film as components, owing to which GR noise is negligible compared to the other two components. It is approximately two orders of magnitude lower than the readout noise.

\section{Characterization of MKIDs}

\subsection{Cryostat and readout system}

We characterized the MKIDs at a temperature below 0.35~K condition.
Our cryostat comprised a vacuum chamber; and first, second, and third radiation shields.
The first and second shields were cryogenically cooled by using a pulse tube refrigerator (PT407RM, Cryomech Co. Ltd)
and maintained at 40~K and 3~K, respectively.
Within the second shield, geomagnetism effects were maintained below 2 mG using magnetic shields (A4K, Amuneal Co., Ltd) set on the shield surface~\cite{21}.
The third shield was a light-tight copper box,
and was cooled using a $^3$He-sorption refrigerator (Niki Glass Co., Ltd.),
inside which we set the MKIDs.
The Wide-MKIDs and the Narrow-MKIDs
were maintained at temperatures of 0.330~K and 0.348~K, respectively and monitored with a 0.003~K accuracy.

\begin{figure}[b!]
  \centering
  \includegraphics[width = 15cm]{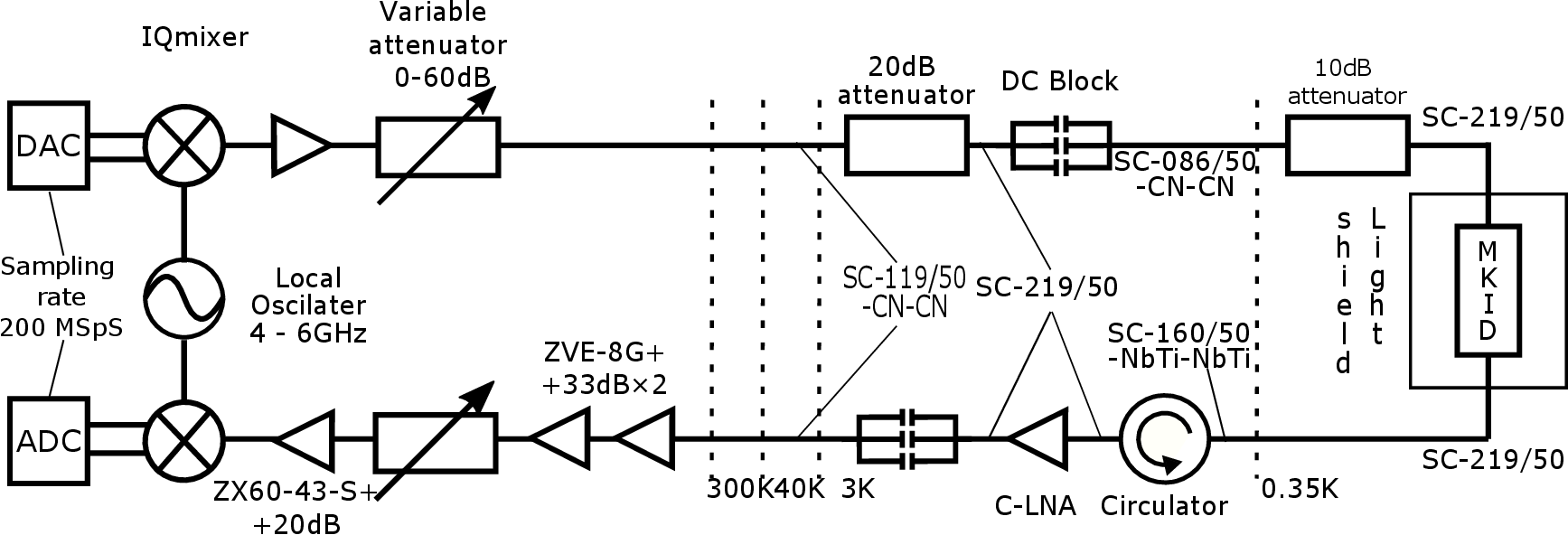}
  \caption{Diagram of the readout system.
  Readout microwaves (bandwidth of 200 MHz) were generated and sampled at a speed of 200 MHz~\cite{23}. 
  The microwaves were up-converted and down-converted using an IQmixer (MLIQ-0218L, Marki Microwave) and a local oscillator (FSL-0010, NI Microwave components).
  The readout power into the MKIDs was amplified with warm amplifiers (ZX60-83LN12+, Mini-Circuit) and tuned with variable attenuators. 
  The readout power out from the MKID was amplified with a cold low-noise amplifier (C-LNA, LNF-LN4\_8C, LOWNOISE FACTORY), as well as warm amplifiers.}
  \label{readouts} 
\end{figure}

A diagram of the readout circuit is shown in Figure~\ref{readouts}.
The microwaves were generated using a digital-to-analog converter~(DAC) with up-conversion using a mixer and a local oscillator (LO).
Additionally, the microwave power at the insertion of the cryostat was adjusted from -19~dBm~to~-81~dBm by using a variable attenuator (LDA-602, Vaunix Co. LTD).
With the attenuation of coaxial cables and attenuators set at 3~K and 0.35~K, respectively, the microwave power into the MKID device ($P_{\rm read}$) was reduced to a range between -60~dBm~and~-122~dBm.
The output microwaves from the MKID were amplified by 40~dB with the C-LNA set to 3~K.
To minimize thermal loading and the attenuation of microwaves, we used a cable of superconducting material to connect the MKID device to the C-LNA.
The output power of the microwaves from the cryostat was maintained in the range of -31~dBm~to~-93~dBm, whereas
the power entering the mixer was set to -11~dBm using the additional variable attenuator.
Return microwaves were measured using an analog-to-digital converter (ADC) after down-conversion by using the mixer and the LO.
Two channels each of the DAC and ADC were employed, at a sampling rate of 200~MHz.
For each frequency tone, the amplitude and phase responses were extracted based on a direct down-conversion logic~\cite{22},
and the down-sampled signals were recorded at a frequency below 1~MHz.

\subsection{Resonant frequencies and quality factors}

\begin{figure}[!h]
  \centering
  \includegraphics[width = 10cm]{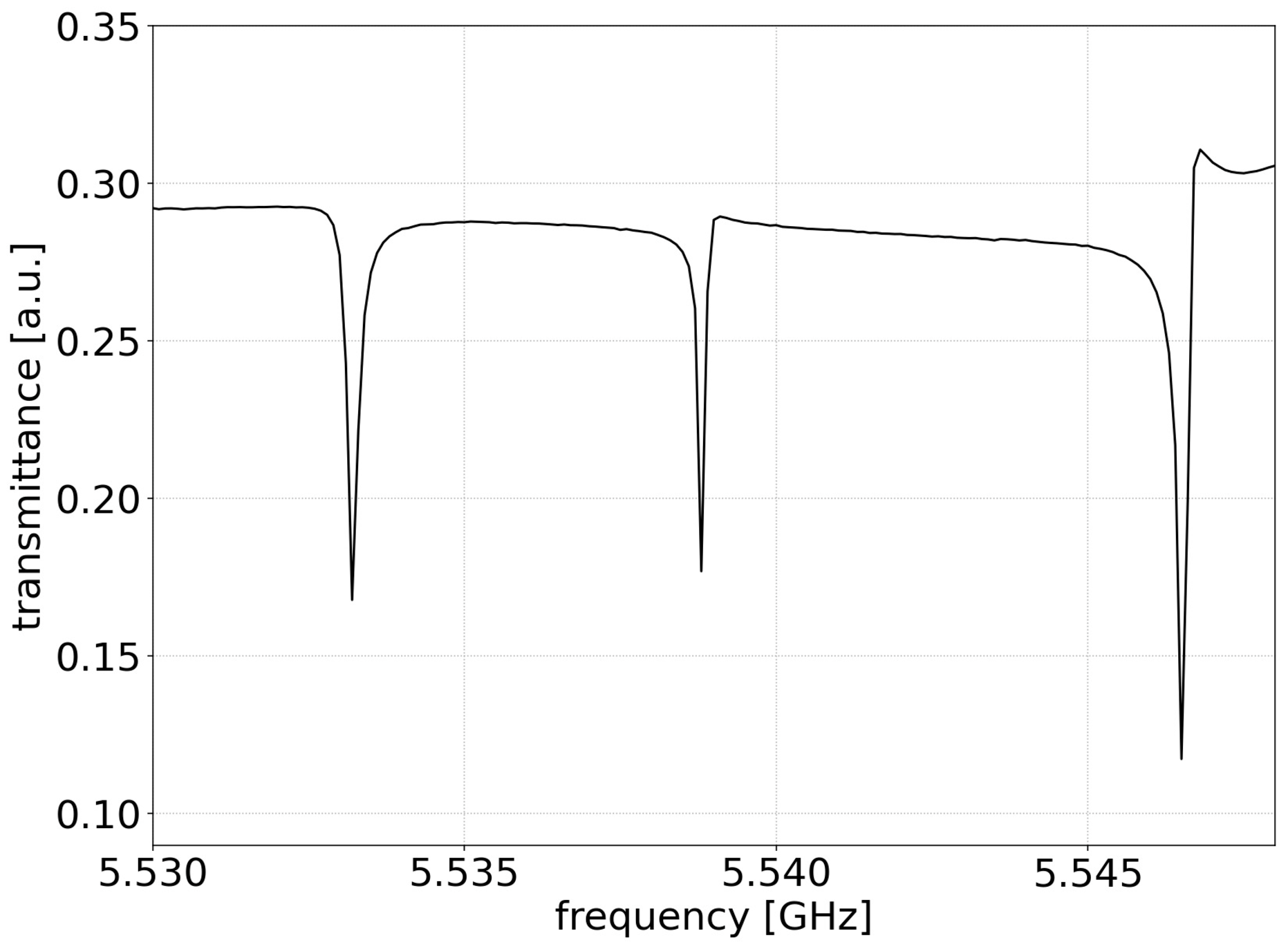}
  \caption{Transmittance of microwaves fed into the Wide-MKIDs.}
  \label{trans_wide}
\end{figure}

\begin{figure}[!h]
  \centering
  \includegraphics[width = 15cm]{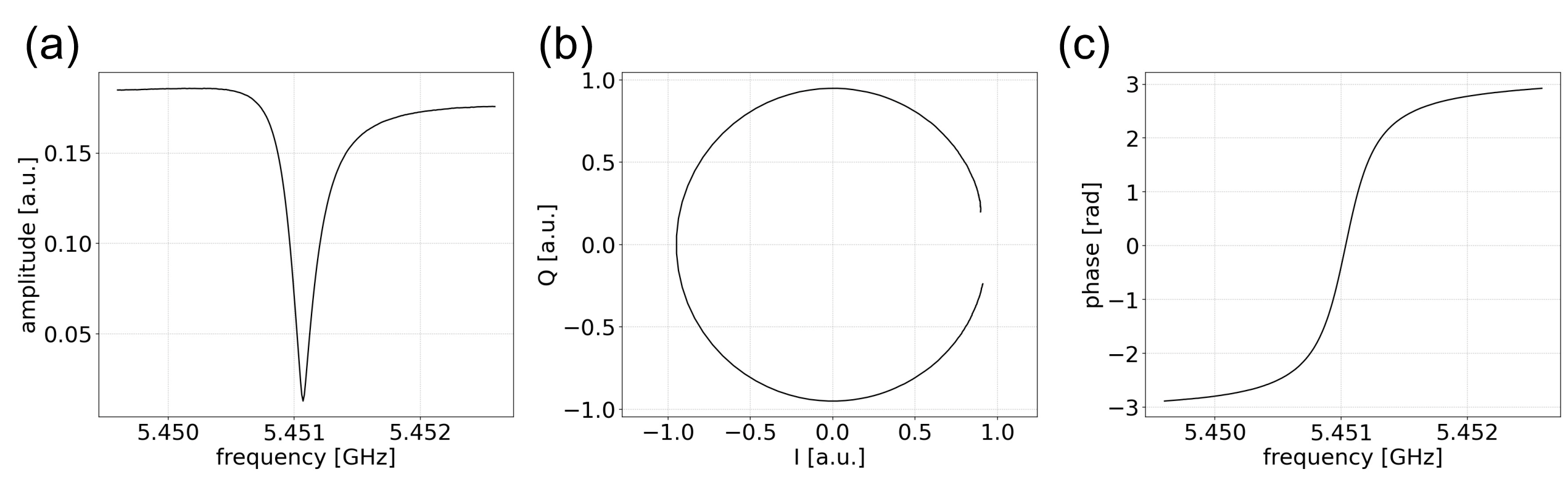}
  \caption{(a) Transmittance as a function of the frequency,
  (b) resonant circle in the $I - Q$ plane, and
  (c) phase of the resonant circle of (b) for one MKID, W1.}
  \label{iq_wide}
\end{figure}

The resonant frequencies were identified based on the transmittance of the microwaves as a function of frequency, as shown in Figure~\ref{trans_wide}.
For each resonance, the in-phase~($I$) and  quadrature~($Q$) components were simultaneously measured, as shown in Figure~\ref{iq_wide}~(b).
Each resonance exhibits a similar circular response in the $I-Q$ plane, and the MKID response is measured by the variation in the radius (amplitude) and phase. 
The $Q_r$ for each MKID is measured as the full width of the half minimum of the transmittance for each resonance.
The radius of the resonant circle represents the ratio $Q_r/Q_c$, and the internal quality factor ($Q_i$) is calculated by using the formula,
\begin{equation}
    {1 \over Q_r} = {1 \over Q_i} + {1 \over Q_c} .
\end{equation}
$Q_i$ represents the energy stored in the resonator over its energy loss.
The resonant frequencies and quality factors for each MKID, measured at $P_{\rm read}$~=~-90~dBm, are summarized in Table~\ref{basic_param}.
To facilitate comprehension, the Wide-MKIDs were labeled as W1, W2, ..., and W8, and the Narrow-MKIDs were labeled as N1, N2, ..., and N9.
Notably, the radiation loss of the microwaves increases with the size of the CPW geometry (details are described in Section 5).
Hence, the $Q_i$ of the Wide-MKIDs is lower than that of the Narrow-MKIDs.

\begin{table}[t!]
  %\caption{Basic parameters of each MKIDs.}
  \caption{Resonant frequencies and quality factors for each MKID.}
  \label{basic_param}
  \centering
  \scalebox{1}{ %%% scale
  \begin{tabular}{c c c c c c}
  \hline \hline
   & fr [GHz] & Qr $\times 10^4$ & Qc $\times 10^4$& Qi $\times 10^4$ \\
  \hline
  N1 & 4.00 & 10.8 & 12.7 & 68.9 \\
  N2 & 4.02 & 10.3 & 11.0 & 77.3 \\
  N3 & 4.03 & 7.32 & 9.15 & 27.2 \\
  N4 & 4.30 & 3.89 & 4.08 & 70.2 \\
  N5 & 4.32 & 3.92 & 4.17 & 63.2 \\
  N6 & 4.32 & 4.02 & 4.23 & 78.6 \\
  N7 & 4.59 & 1.29 & 1.28 & 77.0 \\
  N8 & 4.90 & 1.59 & 1.60 & 73.4 \\
  N9 & 4.92 & 1.64 & 1.55 & 81.0 \\
  \hline
  W1 & 5.45 & 1.52 & 1.60 & 21.0 \\
  W2 & 5.48 & 3.42 & 3.77 & 24.1 \\
  W3 & 5.48 & 1.31 & 1.32 & 40.8 \\
  W4 & 5.49 & 2.82 & 3.81 & 8.57 \\
  W5 & 5.53 & 2.54 & 5.92 & 4.39 \\
  W6 & 5.54 & 5.87 & 10.5 & 12.8 \\
  W7 & 5.55 & 3.51 & 3.57 & 37.6 \\
  W8 & 5.59 & 1.29 & 1.35 & 19.0 \\
  \hline \hline
  \end{tabular}
  }
\end{table}

\subsection{Analysis of power spectrum density}

\begin{figure}[b!]
  \centering
  \includegraphics[width = 10cm]{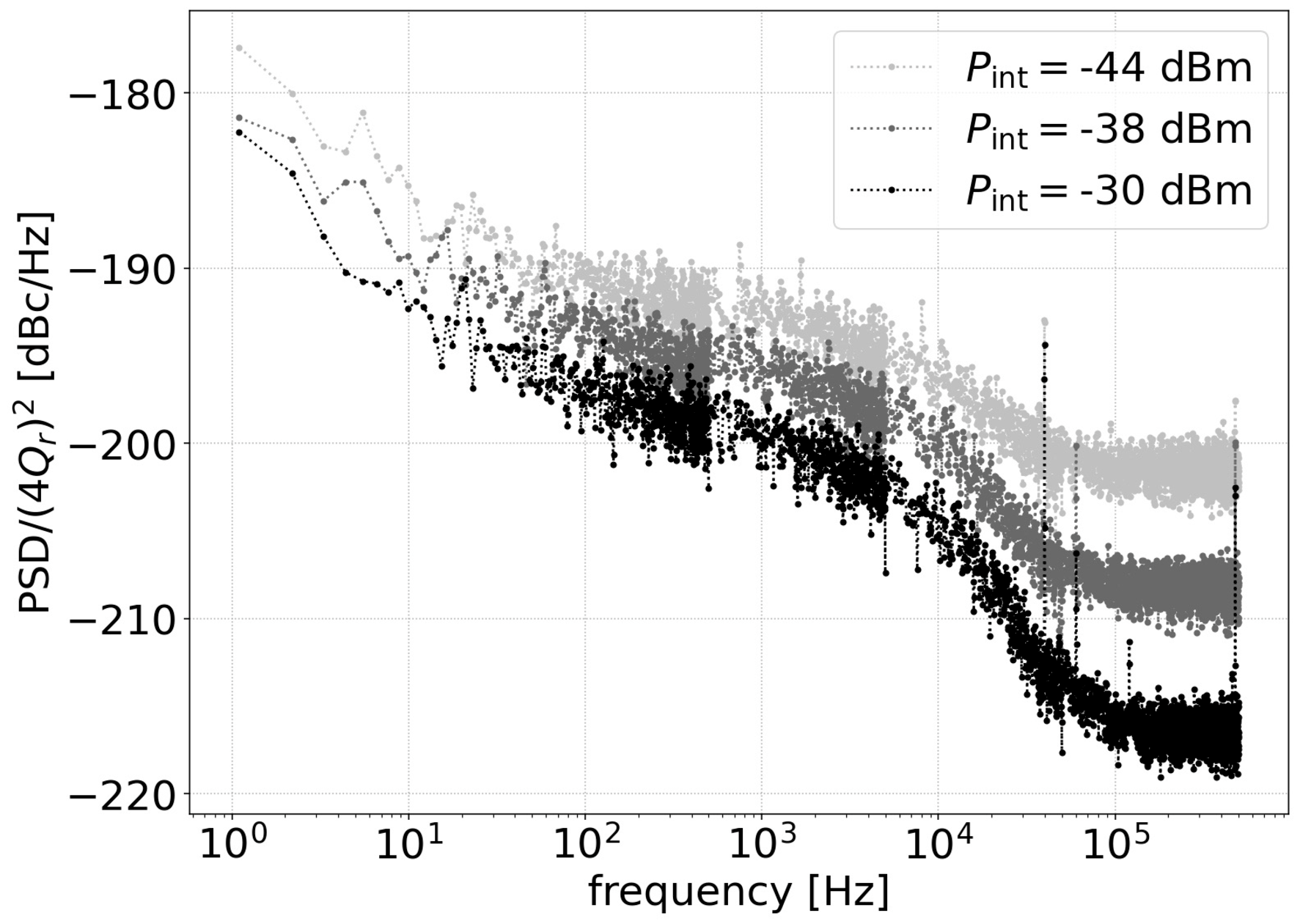}
  \caption{Measured PSDs for the Narrow-MKID (N1) at three different $P_{\rm int}$ conditions.
  The reduction of the normalized PSD was observed while increasing the $P_{\rm int}$.
  As mentioned in Figure~\ref{tls_expect}, there is no response to TLS fluctuation in a short time scale below the resonator ring time ($\sim 10 \mu s$).
  The readout noise ($> 10^5$ Hz) is inversely proportional to the readout power.
  However, the reduction of the PSD in the lower-frequency region ($\leq 10^4$ Hz) is less significant owing to the $P_{\rm int}$ dependence of the TLS noise.
  }
  \label{power_dep_pic}
\end{figure}

The time-ordered data (TOD) for each MKID were measured after identifying the resonant frequency and three TODs were sampled for each MKID at every readout power condition:
100,000 samples with a 1-MHz sampling rate (i.e., duration of 0.1 s),
100,000 samples with a 100-kHz sampling rate (i.e., duration of 1 s),
and 10,000 samples with a 1-kHz sampling rate (i.e., duration of 10 s).
The PSD was obtained as a function of the frequency by applying a fast Fourier transformation~(FFT) for the three TODs.
Figure~\ref{power_dep_pic} presents the PSDs of the Narrow-MKID for three different $P_{\rm int}$ conditions.
Here, the PSD was normalized by the $(4Q_r)^2$ to characterize the TLS noise, as expressed in Eq.~(\ref{eq:PSD_TLS}),
and the $P_{\rm int}$ dependence as well as the frequency dependence was observed.
This is consistent with the estimation in Figure~\ref{tls_expect}.
Moreover, $P_{\rm int}$ was varied from -45~dBm to -30~dBm with 2~dBm step.

The correlation of the TLS noise was characterized using $P_{\rm int}$ and the sampling frequency.
As mentioned above, the contribution of the GR noise was negligible. 
Additionally, the contribution of the readout noise was removed from the final calculation.
The readout noise was estimated as the averages of the high-frequency points ($> 474$ kHz) in each PSD, from which it was subtracted.

\begin{figure}[h!]
  \centering
  \includegraphics[width = 10cm]{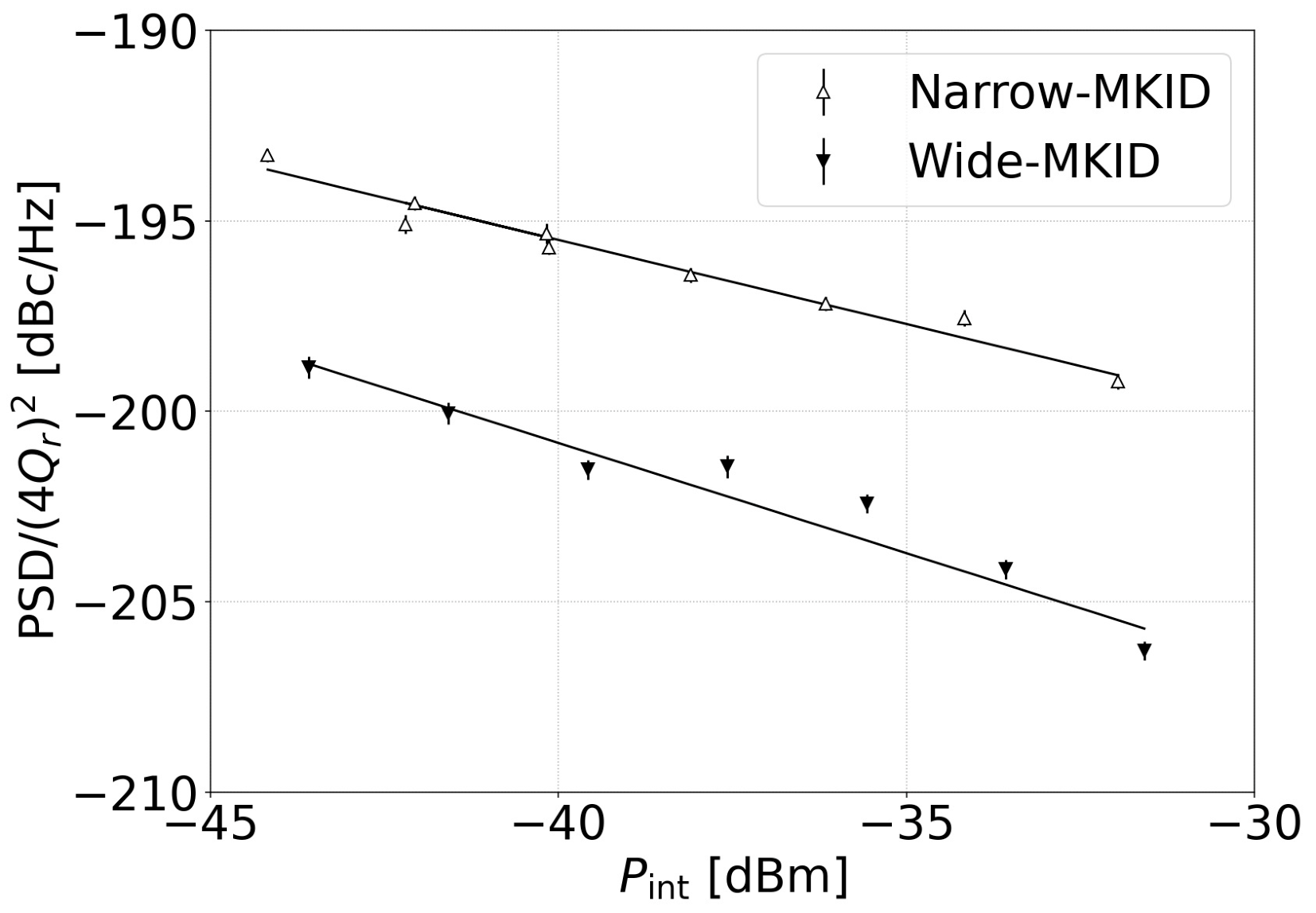}
  \caption{Magnitude of the TLS noise as a function of $P_{\rm int}$}
  \label{power_comp}
\end{figure}

Figure~\ref{power_comp} presents the $P_{\rm int}$ dependence of the TLS noise for one of the Wide-MKIDs and one of the Narrow-MKIDs.
Each data point is an average in the frequency range of 450~--~500~Hz for each $P_{\rm int}$.
The $N_{\rm TLS}$ and $\alpha_P$ were extracted by fitting using the formula Eq.~(\ref{eq:PSD_TLS}). 
The $N_{\rm TLS}$ and $\alpha_P$ values measured for each Narrow-MKID are summarized in Table~\ref{Pint_comp_narrow}, and the measurements for each Wide-MKID are summarized in Table~\ref{Pint_comp_wide}.
According to their averages, the Wide-MKIDs had approximately 5.7~dB lower $N_{\rm TLS}$ than the Narrow-MKIDs.
All the $\alpha_P$ values that were obtained are consistent with our estimation, as discussed in the section 2. Additionally, they are also consistent with the values obtained in previous studies ($\alpha_P \approx -0.5$)~\cite{20}.

\begin{table}[h!]
\caption{Internal power dependence of TLS noise for the Narrow-MKIDs.
For averages, the error was assigned as the standard deviation of each value.}
\label{Pint_comp_narrow}
\centering
\scalebox{1}{
\begin{tabular}{c c c}
\hline \hline
 & $N_{{\rm TLS}}+200~{\rm [dBc/Hz]}$ & $\alpha_P~\times~10^{-1}$\\
\hline
N1 & $5.59 \pm 0.05$ & $ -4.14\pm 0.12$\\
N2 & $4.90 \pm 0.06$ & $ -4.42\pm 0.17$\\
N3 & $4.52 \pm 0.06$ & $ -4.23\pm 0.13$\\
N4 & $4.80 \pm 0.06$ & $ -3.77\pm 0.12$\\
N5 & $3.93 \pm 0.07$ & $ -4.25\pm 0.13$\\
N6 & $4.66 \pm 0.07$ & $ -4.21\pm 0.16$\\
N7 & $5.68 \pm 0.10$ & $ -4.98\pm 0.19$\\
N8 & $5.73 \pm 0.09$ & $ -5.41\pm 0.21$\\
N9 & $5.36 \pm 0.10$ & $ -4.58\pm 0.22$\\
\hline
average & $5.06^{+ 0.56}_{-0.65}$ & $-4.44 \pm 0.49$\\
\hline \hline
\end{tabular}
}
\caption{Internal power dependence of TLS noise for the Wide-MKIDs.
For averages, the error was assigned as the standard deviation of each value.}
\label{Pint_comp_wide}
\centering
\scalebox{1}{
\begin{tabular}{c c c}
\hline \hline
 & $N_{{\rm TLS}}+200~{\rm [dBc/Hz]}$ & $\alpha_P \times 10^{-1}$\\
\hline
W1 & $0.18 ^{+0.20}_{- 0.21}$ & $-8.47 \pm 0.62$\\
W2 & $-0.97 ^{+0.14}_{- 0.15}$ & $-6.61 \pm 0.26$\\
W3 & $-1.70 ^{+0.22}_{- 0.24}$ & $-7.43 \pm 0.78$\\
W4 & $-1.49 ^{+0.21}_{- 0.22}$ & $-7.72 \pm 0.90$\\
W5 & $-1.45 ^{+0.21}_{- 0.22}$ & $-5.59 \pm 0.68$\\
W6 & $-1.22 \pm 0.11$ & $-4.17 \pm 0.58$\\
W7 & $-0.83 \pm 0.12$ & $-5.79 \pm 0.59$\\ 
W8 & $1.33 ^{+0.24}_{- 0.26}$ & $-5.20 \pm 0.35$\\
\hline
average & $-0.65 ^{+1.03}_{- 1.36}$ & $-6.37 \pm 1.44$\\
\hline \hline
\end{tabular}
}
\end{table}

\begin{figure}[h!]
  \centering
  \includegraphics[width = 10cm]{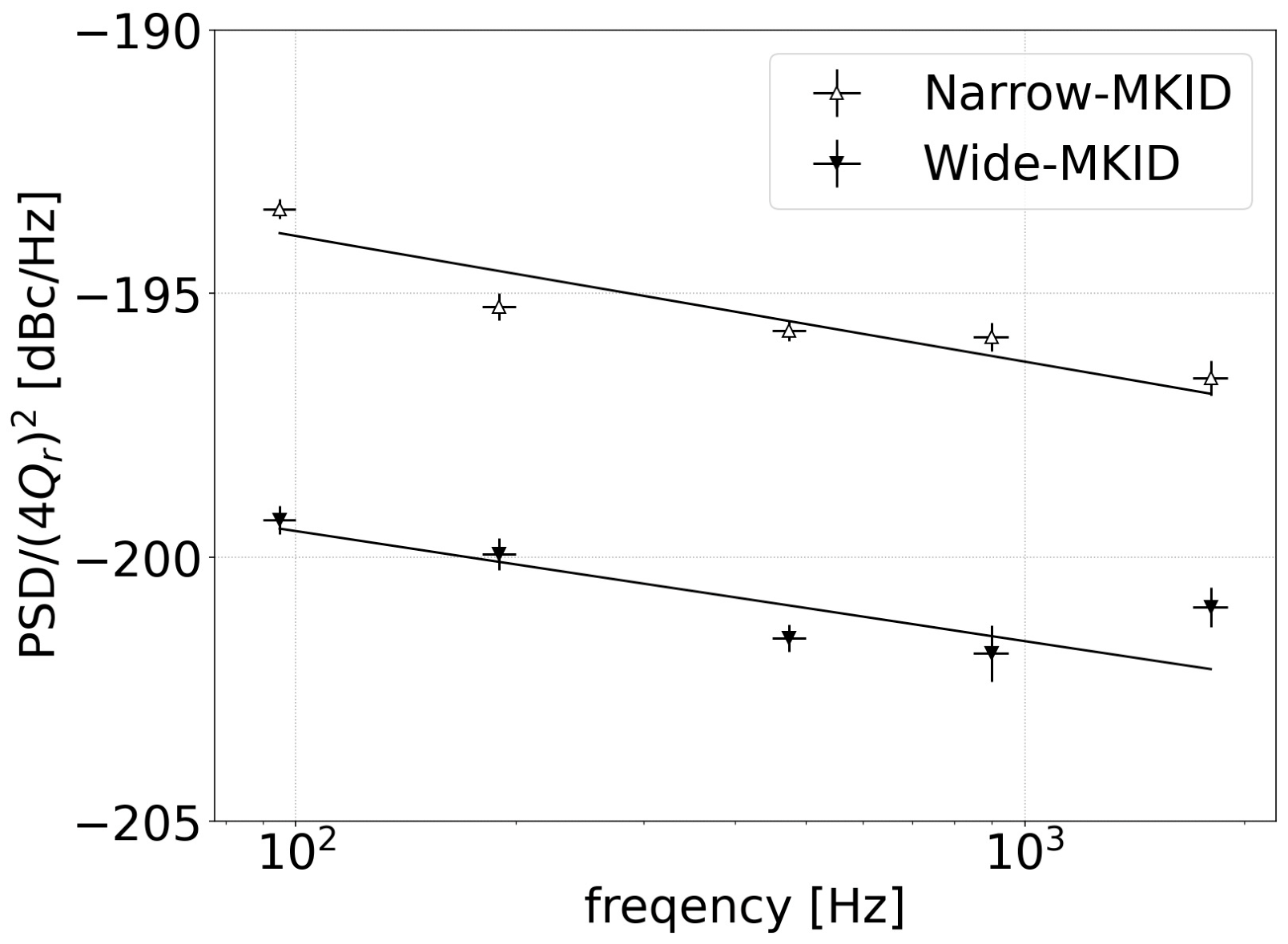}
  \caption{Magnitude of the TLS noise as a function of frequency.}
  \label{freq_comp}
\end{figure}

To characterize the frequency dependence,
averages in the five frequency regions were considered for the data collected under the condition of 
$P_{\rm int} = -40 \pm 1$~dBm:
regions at 90~--~100~Hz, 180~--~200~Hz, 450~--~500~Hz, 900~--~1000~Hz, and 1800~--~2000~Hz.
The results are shown in Figure~\ref{freq_comp}.
The $N_{\rm TLS}$ and $\alpha_f$ were also extracted by fitting using Eq.~(\ref{eq:PSD_TLS}).
The obtained values are summarized in Table~\ref{freq_comp_narrow} and Table~\ref{freq_comp_wide} for the Narrow-MKIDs and the Wide-MKIDs, respectively.
Consequently, it was confirmed that the Wide-MKID has approximately 5.5~dB lower $N_{\rm TLS}$ than the Narrow-MKID, and that $\alpha_f$ is statistically consistent between the Wide-MKIDs and the Narrow-MKIDs.
The overall average ($\alpha_f = -0.24 \pm 0.07$) is also consistent with that of previous studies ($\alpha_f \approx -0.5$), within its uncertainty ($\approx 0.2$)~\cite{19,20}.

\begin{table}[h!]
  \caption{Measured frequency dependence of TLS noise for the Narrow-MKIDs.
  For averages, the error was assigned as the standard deviation of each value.}
  \label{freq_comp_narrow}
  \centering
  \scalebox{1}{
  \begin{tabular}{c c c}
  \hline \hline
   & $N_{{\rm TLS}}+200~{\rm[dBc/Hz]}$ & $\alpha_f \times 10^{-1}$\\
  \hline
  N1 & $5.45 \pm 0.08$ & $-1.80 \pm 0.49$\\
  N2 & $4.82 \pm 0.10$ & $-2.38 \pm 0.61$\\
  N3 & $4.37 \pm 0.10$ & $-3.48 \pm 0.80$\\
  N4 & $4.81 \pm 0.11$ & $-3.42 \pm 1.19$\\
  N5 & $3.81 \pm 0.11$ & $-1.56 \pm 0.63$\\
  N6 & $4.73 \pm 0.10$ & $-2.48 \pm 0.31$\\
  N7 & $5.90 \pm 0.12$ & $-1.80 \pm 1.19$\\
  N8 & $6.03 \pm 0.12$ & $-2.91 \pm 0.80$\\
  N9 & $5.22 \pm 0.13$ & $-2.46 \pm 0.38$\\
  \hline
  average & $5.07 ^{+0.66}_{- 0.78} $ & $-2.46 \pm 0.71$\\
  \hline \hline
  \end{tabular}
  }
  \caption{Measured frequency dependence of TLS noise for the Wide-MKIDs.
  For averages, the error was assigned as the standard deviation of each value.}
  \label{freq_comp_wide}
  \centering
  \scalebox{1}{
  \begin{tabular}{c c c}
  \hline \hline
   & $N_{{\rm TLS}}+200~{\rm [dBc/Hz]}$ & $\alpha_f \times 10^{-1}$\\
  \hline
  W1 & $-0.09^{+0.31}_{-0.33}$ & $-2.54 \pm 0.63$\\
  W2 & $-1.08^{+0.20}_{-0.21}$ & $-1.30 \pm 0.52$\\
  W3 & $-1.01^{+0.23}_{-0.25}$ & $-1.35 \pm 0.52$\\
  W4 & $-0.00^{+0.21}_{-0.22}$ & $-1.79 \pm 0.50$\\
  W5 & $-0.58^{+0.28}_{-0.30}$ & $-2.93 \pm 0.94$\\
  W6 & $-1.14 \pm 0.18$ & $-3.49 \pm 0.54$\\
  W7 & $-0.72 \pm 0.15$ & $-2.08 \pm 0.35$\\
  W8 & $1.91 ^{+0.29}_{-0.31}$ & $-2.58 \pm 0.71$\\
  \hline
  average & $-0.23^{+1.05}_{- 1.38}$ & $-2.26 \pm 0.77$\\
  \hline \hline
  \end{tabular}
  }
  \end{table}

\section{Discussions}
We successfully reduced TLS noise by employing a wide CPW geometry fabricated specifically for this study.
The overall difference in $N_{\rm TLS}$ that we observed between the two types of MKIDs is approximately 5.5 dB, according to the naive averages from Tables 3~--~6.
Uncertainty of $Q_r$ in this measurement is below 2 \%, and it propagates to be 4 \% uncertainty for $N_{\rm TLS}$.
We interpret this improvement results from the reduction in the electric field density at the CPW.
As mentioned in section 3, the magnitude of the electric field density for the Wide-MKID was predicted at approximately 6 dB lower than that for the Narrow-MKID.
As this estimation is close to the observed improvement for the $N_{\rm TLS}$,
it can be inferred that our results support the TLS noise mechanism described in section~2.

However, despite reducing the TLS noise, the wide CPW geometry possesses a disadvantage:
the wide gap in the resonator circuit causes the radiation loss in the microwaves fed into it~\cite{24, 26, 27},
which degrades the internal quality factor,
\begin{equation}
    {1 \over Q_i} = {1 \over Q_{i0}} + {1 \over Q_{\rm loss}}  + {1 \over Q_{\rm load}} ,
\end{equation}
where, $Q_{i0}$, $Q_{\rm loss}$, and $Q_{\rm load}$ represent the quality factors that depend on the material, radiation loss, and external loading, respectively. 
In the laboratory measurement, the external loading was negligible,
and the $Q_i$ was limited by the $Q_{\rm loss}$.
For the Wide-MKID (the Narrow-MKID), we estimate $Q_{\rm loss} \sim {\mathcal{O}(10^4)}~({\mathcal{O}(10^6)})$~
based on numerical simulation~\cite{24,25}.
This estimation explains the differences in $Q_i$ between the two types of MKIDs in Table~\ref{basic_param}. 
In the case of ground-based astronomical observations from the earth,
the effect of the radiation loss is negligible because external loading from the atmosphere is dominant.
For example, to successfully demonstrate the negligible effect of radiation loss, we estimated $Q_{\rm load} \sim 2 \times 10^4$ at the Teide Observatory in the Canary Islands at a frequency band of 150~GHz~\cite{28}.

\section{Conclusions}
We studied TLS noise using two different types of MKIDs.
The magnitude of TLS noise for the MKID with wide CPW geometry was found to be 5.5 dB less than that of that with narrow CPW geometry,
which was close to the expected reduction of 6~dB in the electric field density at the CPW.
Additionally, the indices of the internal power and frequency dependence of the TLS noise are statistically consistent with those of previous studies.
This is the first study to suppress the TLS noise with expanding the CPW geometry using the MKIDs made with the niobium film on the silicon substrate.

\section*{Acknowledgment}
This work was supported by JSPS KAKENHI under grant numbers JP19H05499 and JP20K20427. 
We thank Prof. Koji Ishidoshiro for lending us the cryostat used in this study.
YS also acknowledges JP21J20290.
OT acknowledges support from the Heiwa-Nakajima Foundation, and the US--Japan Science Technology Cooperation Program.
We would like to thank Editage (www.editage.com) for English language editing.

\vspace{0.2cm}

\let\doi\relax

\appendix

\end{document}